\documentclass[twocolumn,showpacs,preprintnumbers,amsmath,amssymb,aps,pre]{revtex4}

\usepackage{mathtools,amssymb,lipsum}

\usepackage{epsfig}

\usepackage{graphicx,latexsym,xcolor}

\usepackage{tikz}

\usepackage[english]{babel}
\usepackage{amsmath}
\usepackage{color}
\usepackage{epsfig}
\usepackage{graphicx}
\usepackage{bm}
\usepackage{times,amsmath,amssymb}
\usepackage{subfigure}
\usepackage{amsfonts}
\usepackage{amssymb}
\usepackage{color}
\global\long\def\bege{\begin{equation}}
\global\long\def\ende{\end{equation}}

\global\long\def\begal{\begin{align}}
\global\long\def\endal{\end{align}}

\begin{document}

\title{Mass generation in graphs}
\author{Ioannis Kleftogiannis$^1$, Ilias Amanatidis$^{2}$}
\affiliation{$^1$ Physics Division, National Center for Theoretical Sciences, Hsinchu 30013, Taiwan }
\affiliation{$^2$Department of Physics, Ben-Gurion University of the Negev, Beer-Sheva 84105, Israel}

\date{\today}
\begin{abstract}
We demonstrate a mechanism for the production of massive
excitations in graphs. We treat the number of neighbors
at each vertex in the graph (degree) as a scalar field.
Then we introduce a mechanism inspired by the Higgs mechanism
in quantum field theory(QFT), that couples the degree
field to a vector-like field, introduced via the graph edges, represented mathematically by the incident matrices
of the graph. The coupling between the two fields produces 
a massless ground state and massive excitations, separated by a mass gap.
The excitations can be treated as emergent massive particles, propagating inside the graph. We study how the size of the graph and its density, represented by the ratio of edges over vertices, affects the mass gap and the localization properties of the massive excitations. We show that the most massive excitations, corresponding to the heaviest emergent particles, localize on regions of the graph with high density, consisting of vertices with a large degree. On the other hand, the least massive excitations, corresponding to the lightest emergent particles
localize on a few vertices but with a smaller degree. Excitations with intermediate masses are less localized, spreading on more vertices instead. Our study shows that emergence of matter-like structures with various mass properties, is possible in discrete physical models, relying only on a few fundamental properties like the connectivity of the models.

\end{abstract}

\maketitle
Random graphs form a good framework for studying
emergent phenomena in physical systems that lack well defined spatio-dimensionality properties or ones that are emergent
\cite{erdos_gallai,aigner,farkas,newman,frieze,berg,mizutaka,paper1,paper2,paper3,bianconi1,bianconi2,gross-yellen,diestel,bollobas}. Such phenomena might underlie mechanisms like the emergence of spacetime
and its properties via discrete models, as alternatives to quantum gravity
approaches\cite{paper1,paper2,paper3,rovelli1,bianconi1,bianconi2,rovelli2,bombelli,fay1,fay2,surya,wolfram,gorard,markopoulou,trugenberger,trugenberger2}. In such discrete models, entities like matter and energy and their attributes, like mass should emerge as manifestations of the connectivity properties of the discrete models, along with spacetime which acts
as the stage for these entities to exist and evolve. Along these lines, we present a mechanism for the emergence of massive excitations in discrete models, that can be applied to any graph based model. We start by assuming that the number of neighbors, connected via single edges at each vertex in the graph, the so called degree of each vertex, forms a scalar field along the graph. Then, inspired by the Higgs mechanism of particle physics\cite{englert,higgs,guralnik,Balaban1984,atlas,djouadi,Dedushenko2023}, we assume that this degree field couples to a vector-like field, generated by introducing directional properties via the graph edges, which can be represented mathematically by the incident matrices of the graph. The Higgs-like mechanism results in the emergence of massive excitations that are separated by a mass gap from a massless ground state. We study how the graph size and density affect the mass-gap along with the structural/ localization properties of the massive excitations.


For our study we consider graphs consisting of $n$ vertices, randomly connected with $m$ edges, where each edge connects only two vertices. There are $\Omega=\binom{ \binom{n}{2}}{m}$ ways to distribute the edges among the vertices. When all the edge configurations have the same probability to appear, $P=\frac{1}{\Omega}$, then we have the case of uniform random graphs $G(V,E)$ where $V$ and $E$ are the sets of vertices $v_i$ and edges $e_i$ respectively\cite{erdos_gallai,aigner,farkas,newman,frieze,berg,mizutaka,paper1,paper2}. Uniform random graphs can be considered as the most generic discrete random structures, constructed by the most minimal number of assumptions. We consider that the density of the graph can be represented by its ratio of edges over vertices $R=\frac{m}{n}$. For $R>0.5$ the graph consists of a large component containing most of the vertices and edges, and many small disconnected components with mostly tree-like structures, whose number decreases as $R$ is increased. Practically
for large $R$, all the small components disappear and only the giant graph component is left.

For our study, we assume that the number of neighbors connected via single edges at each vertex, the degree $d(i)$ at vertex $v_i$, forms a scalar field, spreading along the whole graph. We assume that the degree field has a real value at each vertex 
\begin{equation}
\phi_i=d(i)-\langle d(i)\rangle,
\label{field}
\end{equation}
where $\langle d(i) \rangle$ is the average degree over all the vertices of the large component, for one edge configuration(run) of the graph. At the limiting case of large graphs $n \rightarrow \infty$ we have $\langle d(i)\rangle=2\frac{m}{n}$. The choice of the degree field Eq. \ref{field} ensures that it becomes zero, when the degree is the same (uniform) everywhere on the graph, for example for a complete graph or regular lattices, like the square and the cubic, which should not give rise physically to massive excitations, since there are no field fluctuations. 
For example for a complete graph, there is only one edge configuration where all vertices are connected to each other with $m=\frac{n(n-1)}{2}$ edges.
For this case the degree is the same everywhere $d(i)=n-1=\langle d(i) \rangle$ and the degree field Eq. \ref{field} becomes zero ($\phi_i=0$).
In general we assume that the number of edges for each run of the graph is fixed, giving 
\begin{equation}
\sum_{i=1}^{n} d(i)=2m,
\label{field_con}
\end{equation}
and therefore the degree field satisfies also the following constraint
\begin{equation}
\sum_{i=1}^{n} \phi_i= 0.
\label{field_con}
\end{equation}
Inspired by the Higgs mechanism of particle physics\cite{englert,higgs,guralnik,Balaban1984,djouadi,Dedushenko2023},
we introduce an interaction of the scalar degree
field with a vector-like field defined on the graph,
by using the incidence matrix $B$ of the graph\cite{bianconi1,bianconi2,gross-yellen,diestel,bollobas},
whose elements are given by
\begin{equation}
B_{v_i,e_j} = \begin{cases}
+1 & \text{if $v_i$ is the head of $e_j$},\\
-1 & \text{if $v_i$ is the tail of $e_j$},\\
0 & \text{otherwise}.
\end{cases}
\end{equation}
The resulting incidence matrix $B$ has size $n \times m$.
The head and tail of each edge is chosen by the vertex numbering of the graph. The incidence matrix can be used as a way to introduce directional properties to a graph via its edges. The vector-like field, represented by the directed edges, can be thought as arising from the motion of persistent subgraph structures inside the graph, after applying various evolution mechanisms, for example via external update rules, in emergent spacetime approaches\cite{paper2,paper3,wolfram,gorard}. Although the form of $B$ is dependent on the numbering of the vertices, we have verified that the results of the mass generation mechanism presented in the current manuscript are independent of the particular choice of the vertex numbering. 

The effect of the degree field can be introduced, by assigning a weight to each edge as the sum of the squared degree fields of its two endpoints, the vertices on the head and tail of the edge. This choice ties a Higgs-like field configuration directly to the topology of the graph through its degree distribution. Then a diagonal matrix of the edge weights $W$ can be defined, which has elements, 
\begin{equation}
W_{e_i,e_j} = \begin{cases}
(\phi_{e_{i,tail}})^2+(\phi_{e_{j,head}})^2 & \text{if $i=j$},\\
0 & \text{if $i \neq j$}.\\
\end{cases}
\label{weight_matrix}
\end{equation}
The size of this matrix is $m \times m$.

The coupling between the scalar degree field and the vector-like field represented by the weighted edges can be implemented
by introducing the mass matrix 
\begin{equation}
M = g^2 B W B^{\top}.
\label{mass_matrix}
\end{equation}
which is of size $n \times n$. The constant $g$ represents
the coupling strength between the two fields. For simplicity, we set $g=1$ for of all the calculations presented in the current manuscript. By diagonalizing $M$, which is equivalent to solving the Schrodinger-like equation
\begin{equation}
M \Psi = \lambda \Psi.
\label{mass_equation}
\end{equation}
we get $n$ eigenvalues $\lambda_i$ and their corresponding eigenstates $\Psi^{\lambda_i}$, corresponding to $n$ mass modes. The square root of each eigenvalue $\textit{m}_i=\sqrt{\lambda_i}$ is the mass of each mode. Therefore, the square root of the eigenvalues of $M$ can be considered as the masses of the emergent excitations due to the Higgs-like mechanism. When the weight matrix is the identity matrix $W=I$, the mass matrix $M$ reduces to the Laplacian matrix of the unweighted graph $L=BB^{\top}$.Otherwise Eq. \ref{mass_matrix} is equivalent to the Laplacian of the graph with weighted edges, where the weights are determined by the degree field, as shown in Eq. \ref{weight_matrix}. For systems with the same degree (uniform) on all vertices, for example for regular lattices or a complete graph which has all its vertices connected to each other, we have $d(i)=\langle d(i) \rangle$ leading to $\phi_i=0$ from Eq. \ref{field}. Therefore the weight matrix and the mass matrix become both zero also ($W=0, M=0$). This can be justified physically by the fact that  a uniform field lacks any fluctuations, that could produce field excitations.

\begin{figure}
\begin{center}
\includegraphics[width=0.9\columnwidth,clip=true]{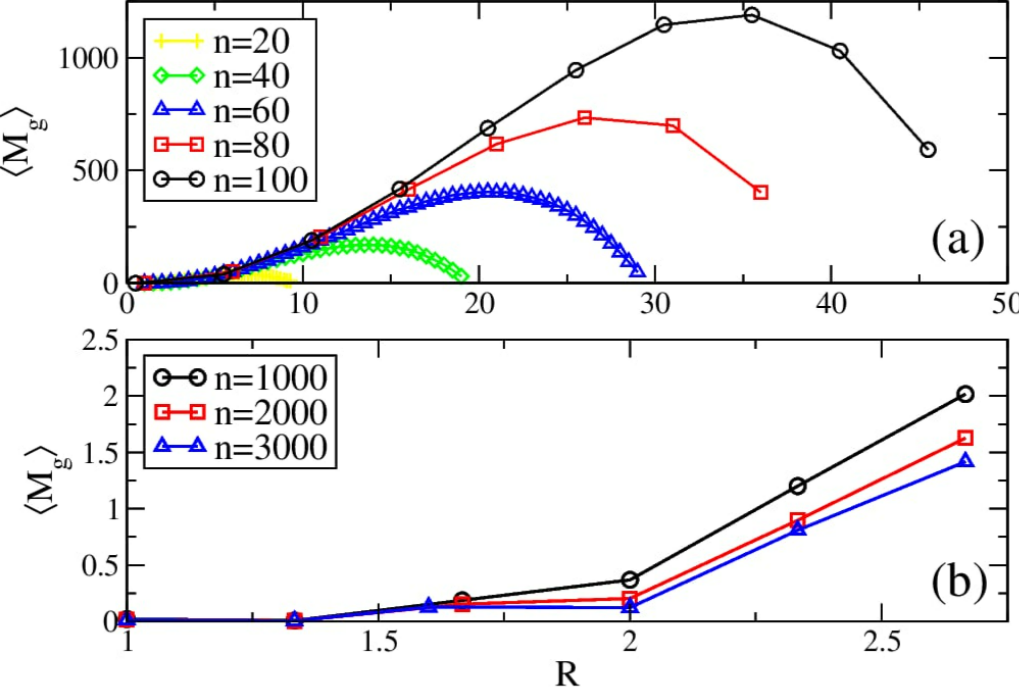}
\end{center}
\caption{a) The average mass gap $\langle M_g \rangle=\langle \lambda_1-\lambda_0 \rangle$, from the massless ground state at eigenvalue $\lambda_0=\textit{m}_0^2=0$, to the first massive excited state with eigenvalue $\lambda_1$, produced by the Higgs-like mechanism on the graph, versus the ratio of edges over vertices $R$ of the graph, representing its density. The different curves are for different graph sizes n. The average of the gap is taken over 1000 configurations(runs) of the graph. The gap initially increases reaching a maximum value and then decreases, as the system approaches the complete graph structure which has all its vertices connected to each other, at ratio $R=\frac{(n-1)}{2}$. For a complete graph we have $M_g=0$ since the degree is the same(uniform) on all the vertices $d(i)=n-1$, and therefore no massive excitations are physically possible. b) The mass gap vs the ratio, for larger graph sizes n and 100 runs, where an increase with $R$ is observed also.}
\label{fig1}
\end{figure}

The mass matrix M is positive semidefinite and therefore all its eigenvalues are non-negative ($\lambda_i \geq 0$). We have found a persistent state at $\lambda_0=0$ corresponding to a massless mode ($\textit{m}_0=0$), which is always the ground state of the system, irrespectively of the graph size n or the ratio R, representing its density. The massless ground state is degenerate for low $R$. As $R$ increases the degeneracy disappears and the wavefunction of the single ground state spreads uniformly on the whole graph, with equal absolute amplitude $|\Psi^{0}_i|=\frac{1}{\sqrt{n}}$ on all the vertices. This state is analogous to the ground state of the Higgs field in particle physics, acting as a constant background energy, assumed to be constant throughout space. The first excited states, for $\lambda_i > 0$, corresponding to the lightest massive excitations, are separated by a mass gap from the ground state at $\lambda_0=0$. In the Higgs picture, these excited states can be thought as the excitations of the Higgs field, corresponding to physically observable massive Higgs particles, detectable in high-energy experiments\cite{atlas}. In Fig. \ref{fig1} we plot the average value of the mass gap $\langle M_g \rangle=\langle \lambda_1-\lambda_0 \rangle$ vs the ratio R representing the graph density, for various graph sizes n. Fig. \ref{fig1}a contains small sizes, with the average taken over 1000 runs, while Fig. \ref{fig1}b contains large sizes for 100 runs. The gap increases initially as the graph becomes denser by increasing $R$. Then as the system approaches the complete graph structure, the gap starts decreasing until it reaches zero. Since a complete graph has all its vertices connecting to each other, the degree is the same for all vertices $d(i)=n-1=\langle d(i) \rangle$, and both the weight matrix and the mass matrix become zero ($W=0$,$M=0$), according to the definitions Eq. \ref{weight_matrix}- \ref{mass_matrix}, resulting in no mass excitations and a zero mass gap.

The form of the massive excitations on the graph and their localization properties can be investigated by calculating the inverse participation ratio (IPR) of the eigenstates 
\begin{equation}
IPR(\lambda)=\sum_{i=1}^{n} |\Psi^{\lambda}_i|^4,
\label{ipr}
\end{equation}
where $\Psi^{\lambda}_i$ is the eigenstate amplitude at vertex $v_i$ of the graph, corresponding to the eigenvalue $\lambda$. Since the eigenstate of the ground state at $\lambda=0$ is equal on all the graph vertices, with amplitude $|\Psi^{\lambda}_i|=\frac{1}{\sqrt{n}}$, we have $IPR(0)=\frac{1}{n}$ for this state. Localized eigenstates, on one vertex for example, have $IPR=1$ instead.  

\begin{figure}
\begin{center}
\includegraphics[width=0.9\columnwidth,clip=true]{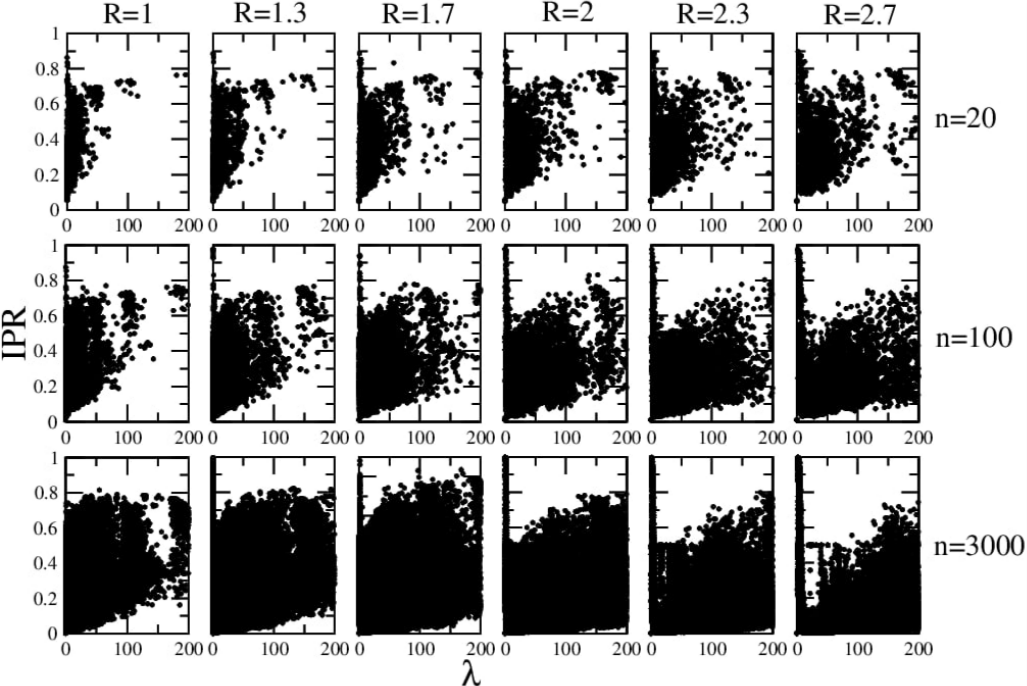}
\end{center}
\caption{The distribution of the inverse participation ratio (IPR) of the eigenstates of the different mass modes, generated by the Higgs-like mechanism on the graph. The IPR is plotted across the eigenvalues $\lambda$ of the mass spectrum. The different panels correspond to various graph sizes $n$, running across the panel rows, and ratios of edges over vertices $R$, representing the graph density, running across the panel columns. A peak of the IPR is gradually formed near $\lambda=0$, as $R$ increases, for all n, meaning that the first excited mass modes, corresponding to the lightest emergent particles, become localized on a few vertices of the graph. As the graph size $n$ increases, all the mass modes become less localized in average, indicated by the decreasing values of IPR. Inside each panel, the states become in average more localized as the mass increases.}
\label{fig2}
\end{figure}

In Fig. \ref{fig2} we show various panels with the distribution of $IPR$ for states across the mass spectrum, for various graph sizes n and densities $R$. A peak near $\lambda=0$ is formed gradually as the graph becomes denser by increasing $R$, for all $n$, indicating that the corresponding massive excitations, the ones with the smallest masses(lightest particles), become localized on a few vertices of the graph. The rest of the  states, away from $\lambda=0$, become in average less localized as the graph becomes denser, by increasing $R$, indicated by the decreasing values of $IPR$.

\begin{figure}
\begin{center}
\includegraphics[width=0.9\columnwidth,clip=true]{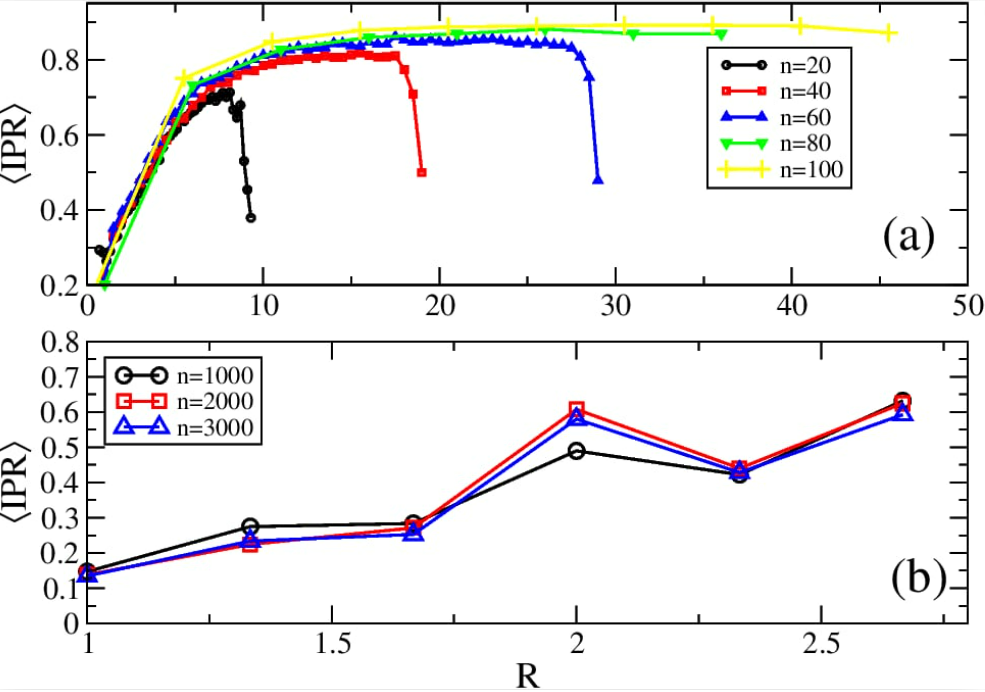}
\end{center}
\caption{ a) The average value of the inverse participation ratio $\langle IPR \rangle$ for eigenstates of the first excited massive modes, for various graph sizes $n$, and 1000 runs. The $\langle IPR \rangle$ is plotted vs the ratio of edges over vertices $R$. b) The $\langle IPR \rangle$ for larger graph sizes and 100 runs. In both panels the $\langle IPR \rangle$ increases with $R$, unless the system lies near the complete graph case, indicating that the first excited massive modes become more localized, as the graph density increases.
}
\label{fig3}
\end{figure}

In Fig. \ref{fig3} we plot the average value $\langle IPR \rangle$ versus $R$ for the first excited massive state with eigenvalue $\lambda_1$, right above the massless mode with eigenvalue $\lambda_0=0$. Results for small(large) sizes and 1000(100) runs are shown in Fig. \ref{fig3}a(b). The states becomes less localized as $R$ increases and the graph becomes denser, unless the system lies near the complete graph case, where all vertices are connected to each other.

\begin{figure}
\begin{center}
\includegraphics[width=0.9\columnwidth,clip=true]{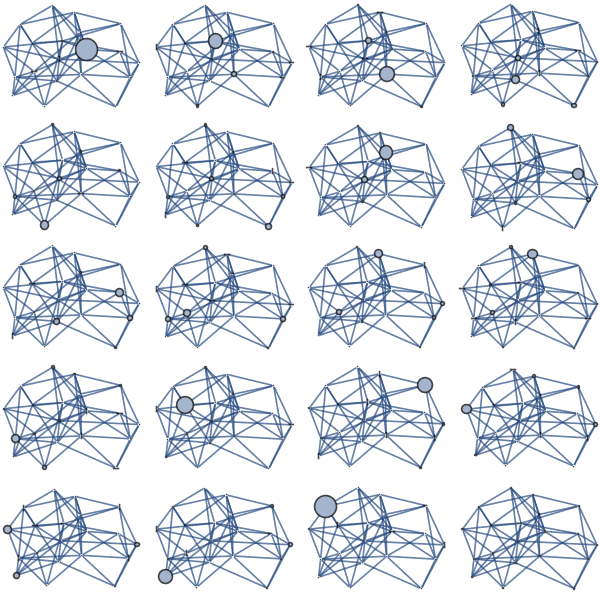}
\end{center}
\caption{Several schematics demonstrating the eigenstates of all the mass modes (eigenvalues), due to the Higgs-like mechanism on a graph of size $n=20$ with ratio of edges over vertices $R=3 (m=60)$. We have plotted only the largest(giant) component of the graph. The eigenstate probability $|\Psi^{\lambda}_i|^2$, corresponding to eigenvalue $\lambda$ is represented by the radius of the circles on each vertex $v_i$, tuned by a scale factor. The panels are arranged from larger to smaller $\lambda$, so that the upper-left(lower-right) schematic corresponds to the largest(smallest) mass. The heaviest and lightest masses are localized on a few vertices of the graph, while the the intermediate masses spread over the whole graph instead. Also the heaviest masses are localized on vertices with a larger degree than the lightest masses.}
\label{fig4}
\end{figure}
\begin{figure}
\begin{center}
\includegraphics[width=0.9\columnwidth,clip=true]{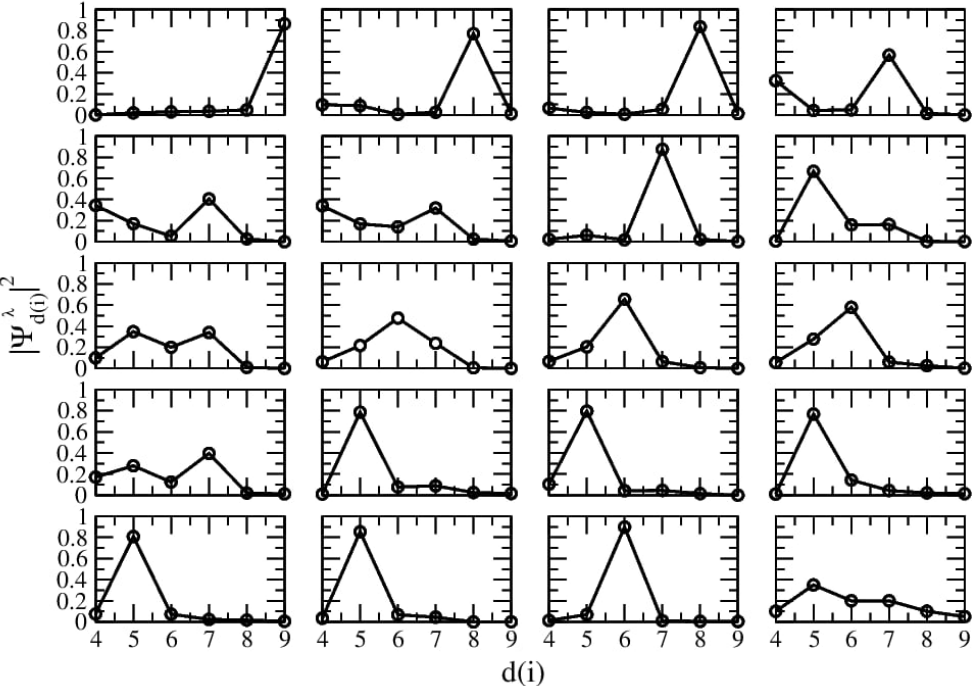}
\end{center}
\caption{The eigenstate probability $|\Psi^{\lambda}_{d(i)}|^2$, for eigenvalue $\lambda$ and vertex degree $d(i)$, of all the mass modes for the giant component of a graph of size $n=20$ and ratio of edges vs vertices $R=3(m=60)$. The panels are arranged from larger to smaller $\lambda$, so that the upper-left(lower-right) schematic corresponds to the largest(smallest) mass. The heaviest masses are concentrated in average on vertices with a larger degree than the lightest masses.}
\label{fig5}
\end{figure}

In Fig. \ref{fig4} we show the graph structure for $n=20$ and $R=3(m=60)$ along with the eigenstate probability $|\Psi^{\lambda}_i|^2$, whose magnitude is represented by the radius of the filled circles, tuned by a scale factor.
The schematics are arranged from larger to smaller $\lambda$,
meaning that the upper-left(lower-right) schematic corresponds to the largest(smallest) mass. The eigenstates corresponding to the largest and smallest masses (schematics in the first and third/fourth rows of Fig. \ref{fig4}),  are more localized on the graph, than eigenstates corresponding to intermediate masses. Therefore, the lightest and heaviest emergent particles are more localized than the particles with intermediate masses. We observe also that the heaviest particles, for example the cases shown in the first row of Fig. \ref{fig4}, are localized on vertices with high edge connectivity (degree). This can be seen in Fig. \ref{fig5}, where we plot the probability $|\Psi^{\lambda}_{d(i)}|^2$ for each degree. Eigenstates at the upper end of the mass spectrum, with the highest masses have a larger probability to appear on vertices with a large degree. If the local graph density is also large, this leads naturally to a large Ollivier-Ricci graph curvature,
which converges to the Riemannian curvature of manifolds, when a lot of vertices are considered. Along these lines, it would be interesting in future studies to explore the connection of this effect, to the general relativistic effect of large concentration of mass-energy, causing a large spacetime curvature.

We have to note also that we can define a mass matrix for the edges of the graph as, $M_1 = g^2 B^{\top} W_1 B$, which has size $m \times m$. The weight matrix $W_1$ can be defined as a diagonal matrix, written in the basis of vertices, with each diagonal element being simply the degree field squared, $\phi^2_i$, at vertex $v_i$. For $W_1=I$ the edge mass matrix $M_1$ reduces to the edge Laplacian matrix of the unweighted graph $L_1=B^{\top}B$. It would be interesting in future studies to investigate the mass-gap properties for the eigenvalues of $M_1$, along with the localization properties of its eigenstates. The modes of $M_1$ could represent the emergent mass modes(excitations) of the vector-like field, defined via the incident matrices of the graph.

To conclude, we have presented a Higgs-like mechanism for the generation of mass in graphs. The mechanism couples a degree field defined on each vertex of the graph, according to its number of neighbors, to a vector-like field defined via the edges of the graph. We have shown that this mechanism leads
to the emergence of massive excitations with various interesting localization properties. The massive excitations are separated by a large mass gap from a massless ground state, which spreads uniformly on the whole graph. The massive excitations can be treated as emergent particles with various masses, propagating inside the graph. We have found that as the graph becomes denser, the mass gap increases and the massive excited states become in average gradually more localized across the graph. In addition the heaviest particles, in the upper end of the mass spectrum are localized on a few vertices, which have a large degree. The lightest particles, corresponding to the first excited massive states are also localized on a few vertices of the graph but with a smaller degree. Intermediate masses are more evenly distributed and less localized, spreading on more vertices instead. Our study shows that mass generation is possible in discrete physical models, by considering a minimal mechanism based only on the graph connectivity properties, without any additional assumptions like extra fields that live on top of the models. This mechanism might therefore drive mass generation in the most minimal discrete models of emergent spacetime.


\section*{References}


\begin{thebibliography}{99}

\bibitem{erdos_gallai}P. Erd\H{o}s, T. Gallai,  "Gráfokel\H{o}írt fokszámú pontokkal", Matematikai Lapok, {\bf 11}: 264-274 (1960).

\bibitem{aigner}M. Aigner, E. Triesch, Discrete Math.{\bf  136}, 3-20 (1994).

\bibitem{farkas}I. J. Farkas, I. Der\'{e}nyi, A.-L. Barab\'{a}si, and T. Vicsek,  Phys. Rev. E {\bf 64}, 026704 (2001).

\bibitem{newman}M. E. J. Newman  Networks: An Introduction Oxford University Press, Oxford (2010).

\bibitem{frieze}A. Frieze, M. Karonski, Introduction to random graphs. Cambridge University Press (2015).

\bibitem{berg}J. Berg and M. Lassig., Phys. Rev. Lett. {\bf 89}, 228701 (2002).

\bibitem{mizutaka}S. Mizutaka and T. Hasegawa, Journal of Physics: Complexity {\bf 1}, 035007 (2020).

\bibitem{paper1}I. Kleftogiannis and I. Amanatidis,
Phys. Rev. E {\bf 105}, 024141 (2022).

\bibitem{paper2}I. Kleftogiannis and I. Amanatidis, 
arXiv:2210.00963 [cond-mat.dis-nn] (2022).

\bibitem{paper3}I. Kleftogiannis and I. Amanatidis, 
arXiv:2507.01468 [cond-mat.dis-nn](2024).
\bibitem{bianconi1} G. J. Bianconi, J. phys. Complex. {\bf 2}, 035022 (2021).

\bibitem{bianconi2} G. J. Bianconi, J. Phys. A: Math. Theor. {\bf 57}, 015001 (2023).


\bibitem{gross-yellen} Jonathan L Gross, Jay Yellen, Graph Theory and its applications, second edition, 2006 (p 97, Incidence Matrices for undirected graphs; p 98, incidence matrices for digraphs).

\bibitem{diestel} Diestel, Reinhard (2005), Graph Theory, Graduate Texts in Mathematics, vol. 173 (3rd ed.), Springer-Verlag, ISBN 3-540-26183-4.

\bibitem{bollobas} Bollobás B. 1998. In Modern Graph Theory, ed. S Axler, FW Gehring, KA Ribet, pp. 215–52. New York:
Springer Sci. Bus. Media.

\bibitem{rovelli1} C. Rovelli, Class. Quantum Grav. {\bf 28}, 153002 (2011).
\bibitem{rovelli2} C. Rovelli,  Living Rev. Relativ. {\bf 11}, 5 (2008).
\bibitem{bombelli}L. Bombelli, J. Lee,  D. Meyer and R. D. Sorkin, Phys. Rev. Lett.  {\bf 59},   521-524 (1987).
\bibitem{fay1}F. Dowker, Annals of the New York Academy of Sciences {\bf 1326}, 18-25, (2014).
\bibitem{fay2}F. Dowker and S. Zalel,   C R Physique {\bf 1}, 8:246-253 (2017).
\bibitem{surya}S. Surya,  Living Rev. Rel. {\bf 22},  5 (2019).

\bibitem{wolfram}S. Wolfram, Complex Systems {\bf 29}, (2)   pp. 107-536 (2020).
\bibitem{gorard}J. Gorard, Complex Systems, {\bf 29}, (2)  pp. 599-654 (2020).
\bibitem{markopoulou}T. Konopka, F. Markopoulou and S. Severini, Phys. Rev. D {\bf 77}, 104029 (2008).

\bibitem{trugenberger}C.A. Trugenberger, J. High Energ. Phys.  45 (2017).
\bibitem{trugenberger2}C. Kelly, C. Trugenberger, and F. Biancalana, Class. Quantum Grav. {\bf 38}, 075008 (2021).


\bibitem{englert} F. Englert and R. Brout, Phys. Rev. Lett. \textbf{13}, 321 (1964).

\bibitem{higgs} P. W. Higgs, Phys. Rev. Lett. \textbf{13}, 508 (1964).

\bibitem{guralnik} G. S. Guralnik, C. R. Hagen, and T. W. Kibble, Phys. Rev. Lett. \textbf{13}, 585 (1964). 

\bibitem{Balaban1984}
T. Balaban, J. Imbrie, A. M. Jaffe, and D. Brydges,
Annals of Physics \textbf{158} 281 (1984).

\bibitem{atlas} ATLAS Collaboration, Phys. Lett. B \textbf{716}, 1 (2012).

\bibitem{djouadi} Abdelhak Djouadi, AIP Conf. Proc. \textbf{1444}, 45–57 (2012). 

\bibitem{Dedushenko2023}
M. Dedushenko, International Journal of Modern Physics A \textbf{38} 04n05 (2023).




\end{thebibliography}
\end{document}